\documentclass[11pt,a4paper]{article}
\usepackage[english]{babel}
\usepackage[utf8]{inputenc}
\usepackage{graphicx}
\usepackage{xcolor}
\usepackage{epstopdf}
\usepackage{pdfpages}
\usepackage{hyperref, url, footnote}
\usepackage{amsmath,amsfonts,amssymb,amsthm,nccmath,latexsym,mathtools, cases, cite}
\usepackage[hmargin=3cm,vmargin=3cm,bmargin=3cm]{geometry}
\usepackage{multirow}
\usepackage{color}

\def\ni{\noindent}

\def\vs{\vspace{0.5cm}}

\begin{document}
\centerline{\LARGE Average Thermal Evolution of the Universe}
\vs

\centerline{Natacha Leite$^{1,2}$ and Alex H. Blin$^1$}
\vs

\ni{\sl $^1$CFC, Departamento de Física, Universidade de Coimbra, 3004-516 Coimbra, Portugal}
\vs

\ni{\sl $^2$II. Institut für Theoretische Physik, Universität Hamburg, Luruper Chaussee 149,\\
22761 Hamburg, Germany}

\begin{abstract}
We sketch the average behaviour of the temperatures and densities of the main components of the $\Lambda$CDM universe
after inflation. It is modelled as a perfect fluid with dark energy associated with the macroscopic effect of conformal variations of the metric.
The main events of the thermal evolution are studied, such as the effect of particle annihilations and decoupling, and the transitions between the eras dominated
by different entities. Estimates of the average present epoch temperature of baryonic matter and dark matter composed of neutralinos are given.
We study the eventual presence of a sterile neutrino component and find that the sterile neutrino density at the epoch of primordial nucleosynthesis
is in agreement with expectations when their evolution starts, at the end of inflation, in temperature equilibrium with the rest of the universe.
\end{abstract}

\section{Introduction}
The aim of this paper is to study the evolution of the scale factor and the global behaviour of
temperatures and densities of the components of the universe after the end of the epoch of inflation and how these quantities are influenced by the property of the neutrinos being Majorana or Dirac particles and by the choice of dark matter candidate masses.
In this description of the average evolution neither the details of the primordial nucleosynthesis nor the
impact of structure formation on the matter temperature are taken into consideration. We assume dark matter
to be of only one type and presume that the chemical potentials are negligible in the regimes we
are interested in.\\

To track the thermal history of the universe, let us consider the standard cosmological picture based on the observational evidence that the universe is highly homogeneous, isotropic and flat, and composed of approximately 72\% dark energy, 23\% dark matter and 5\% baryonic matter \cite{WMAP7}.\\
Near the end of the twentieth century, astronomical observations of redshifts of supernovae  \cite{perlmutter, riess} have led to the conclusion that the
universe is expanding at an accelerated rate. This behaviour indicates the presence of an entity with negative pressure, the dark energy, and the concept of a cosmological constant in Eintein's equations has thus been revived. It acts as a repulsive gravity responsible for the accelerated expansion of the universe.\\
One way to account for the origin and nature of the cosmological constant is by considering quantum fluctuations of the metric \cite{narlikar, Blin}. At the Planck scale, gravity and quantum mechanics can not be seen as independent. Space-time is expected to be coarse grained, requiring a quantum metric tensor.\\
The simplest way of generalizing the metric to include quantum effects is through a scalar field $\varphi$ describing conformal variations of the metric around
its classical value. The advantage of conformal variations is that causality is obeyed, since the light cone structure remains intact. These fluctuations can be written as follows \cite{narlikar, Blin}
\begin{equation}
g_{\mu \nu} = (1+\varphi)^2\bar{g}_{\mu \nu}=\Phi^2\bar{g}_{\mu \nu},
\end{equation}
where $\bar{g}_{\mu \nu}$ represents the usual classical metric tensor about which the fluctuations occur. The fluctuation average is required to be $<\varphi>=<\varphi_{, \mu}>=0$, centring the generalized metric around its classical value and yielding no drift of $\varphi$ in space-time.\\
The presence of fluctuations changes the Ricci tensor $R_{\mu \nu}$ and curvature scalar $R$, and when deducing Einstein's equations from variation
of the Einstein-Hilbert action with respect to the classical metric {\sl and} the fluctuation field, one gets (units $\hbar=c=1$)\cite{Blin}
\begin{equation} \label{eq:maisuma}
\bar{R}_{\lambda \mu} - \frac{1}{2}\bar{g}^{\lambda \mu}\bar{R} - \bar{g}_{\lambda \mu}\Lambda= 8 \pi G T_{\lambda \mu},
\end{equation}
with
\begin{equation} \label{eq:Lambda}
\Lambda = -\frac{1}{4}\left(\bar{R} + 8\pi G T_{\lambda \mu}\bar{g}^{\lambda \mu}\right),
\end{equation}
where the quantities with a bar refer to the classical expressions without fluctuations. Equation (\ref{eq:maisuma}) is Einstein's equation with a cosmological constant (\ref{eq:Lambda}) that stems from the fluctuations.
It is possible to show \cite{Blin} that the magnitude of $\Lambda$ is consistent with the observational values attributed to dark energy. Thus, the accelerated expansion that dark energy accounts for in standard cosmology can be seen as a consequence of considering quantum fluctuations present at Planck scale. \\
For the universe as a perfect fluid, the stress-energy tensor has the form
\begin{equation} \label{eq:Tmunu}
T^{\mu \nu}  = \left(\rho + P\right)u^\mu u^\nu - Pg^{\mu \nu},
\end{equation}
where $\rho$ represents the mass-energy density and $P$ represents the pressure of the fluid. We can use it in equation (\ref{eq:Lambda}) to ascertain that the cosmological constant derived this way is indeed constant. Taking the time derivative of equation (\ref{eq:Lambda}), we confirm \cite{tese}
that $\dot{\Lambda} =0$.  \\
We will use the Robertson-Walker metric, appropriate for a homogeneous and isotropic universe, as observations seem to indicate ours is,
\begin{equation}
ds^2=dt^2-Q(t)^2\left[\frac{dr^2}{1-kr^2}+r^2(d\theta^2+\sin^2\theta d\phi^2)\right],
\end{equation}
where $Q(t)$ represents the scale factor and $k$ the curvature of the space.\\
The standard model for cosmology stipulates that the universe originated from a hot dense state, that expanded and cooled down with time.
This Big Bang model can be complemented by inflationary theories, which propose a brief period of exponential  growth for early times.
We can study what happened after inflation, by using Friedmann's equation
\begin{equation} \label{eq:00}
3\frac{\dot{Q}^2 + k}{Q^2} - \Lambda = 8\pi G \rho.
\end{equation}
It is customary to divide the density by the critical density ($H_0$ is the Hubble constant)
\begin{equation} \label{eq:rho_c}
\rho_c=\frac{3H_0^2}{8 \pi G},
\end{equation}
and work with the density parameter $\Omega=\rho/\rho_c$.
We can define different density parameters for the different entities, namely the density parameter for matter,   $\Omega_{m}$ (where we can further
distinguish between the density of baryonic matter $\Omega_{b}$ and cold dark matter $\Omega_{c}$) and for radiation,
$\Omega_{r}$. Similarly, the cosmological constant and curvature can be expressed by the density parameters  $\Omega_{\Lambda}=\Lambda/3H_0^2$ and
$\Omega_{k}= -k/(Q(t)^2H_0^2)$.\\
It will also be important to consider the equation that describes the energy-mass conservation
\begin{equation} \label{eq:rhoponto}
\dot{\rho}_i = -3\frac{\dot{Q}}{Q}(\rho_i + P_i),
\end{equation}
which can be written independently for each component $i$, as long as there is no significant conversion between them. \\
We can use equations of state \cite{KeT}
\begin{equation}
\omega_i = \frac{P_i}{\rho_i},
\end{equation}
to integrate the energy conservation equation, for convenience written in the form
\begin{equation} \label{eq:dotomega}
\frac{\dot{\Omega_i}}{\Omega} = -3(1+\omega_i)\frac{\dot Q}{Q},
\end{equation}
and relate the density parameter of each entity with the scale factor.
This leads to the following behaviour. For radiation
\begin{equation} \label{eq:3rad}
\omega_r = \frac{1}{3},\hspace{1 cm} \Omega_r \propto Q(t)^{-4},\hspace{1 cm} T_r \propto Q(t)^{-1};
\end{equation}
for matter (dust)
\begin{equation} \label{eq:mat3}
\omega_m = 0,\hspace{1 cm} \Omega_m \propto Q(t)^{-3},\hspace{1 cm} T_m \propto Q(t)^{-2};
\end{equation}
and for dark energy
\begin{equation}
\omega_\Lambda = -1, \hspace{1 cm} \Omega_\Lambda \propto \text{constant}.
\end{equation}

\section{Simulation procedure}

\subsection{Modelling the density}

The proportionality between the density and the temperature of the matter contributions is easy to find knowing the present time ($t_0$) value of $\Omega_m$
and of $Q(t)$ (which we choose to set as $Q(t_0)=1$), but not in the radiation case. In the early universe, when the temperature was high enough, other particles
were relativistic and were in thermal equilibrium with photons, being part of radiation.
We can address this by counting the effective degrees of freedom of each species, to obtain $N(T)$ for the initial relativistic particle mixture and for
subsequent times.
This is done by considering that a species annihilates when its rest mass is higher than the temperature, the point at which it stops contributing
to $N(T)$. As the temperature decreases with time, different particles annihilate with their antiparticles when the
temperature of the universe reaches their mass threshold, reducing the number of degrees of freedom and liberating thermal energy of annihilation in the process.
This occurs until only photons remain thermalized.
Reading from bottom to top, Table \ref{table:N} lists the values of $N(T)$ as standard model particles are annihilated.
Alternatively, we can
visualise the inverse process: as the temperature increases when we travel backwards in time, new particles are created
when the temperature of the universe
reaches their mass threshold, unfolding new degrees of freedom.\\
\begin{table}[!ht]
\begin{center}
{\begin{tabular}{cc|c|c||c|}
\cline{3-5}
& &  \multicolumn{3}{ c| }{4$N$}  \\ \hline
\multicolumn{1}{ |c| }{$T_r$} & New particles & 4$N_r(T_r)$ & 4$N_\nu(T_\nu)$& 4$N_s(T_s)$\\ \hline
\multicolumn{1}{ |c| }{$T_r < m_e$} & $\gamma$'s  & 8 & \multirow{2}{*}{21} & \multirow{13}{*}{21} \\ \cline{1-3}
\multicolumn{1}{ |c| }{$m_e< T_r < T_{D_\nu}$} & $e^\pm$  & 22 & &  \\  \cline{1-4}
\multicolumn{1}{ |c| }{$T_{D_\nu} < T_r < m_\mu$} & $\nu$'s & \multicolumn{2}{ |c|| }{43} & \\ \cline{1-4}
\multicolumn{1}{ |c| }{$m_\mu< T_r < m_\pi$} & $\mu^\pm$ & \multicolumn{2}{ |c|| }{57} & \\ \cline{1-4}
\multicolumn{1}{ |c| }{$m_\pi< T_r < T_c$} & $\pi$'s & \multicolumn{2}{ |c|| }{69} &  \\ \cline{1-4}
\multicolumn{1}{ |c| }{$T_c< T_r < m_s $} & $u\bar{u}, d\bar{d}, g$'s$-\pi$'s & \multicolumn{2}{ |c|| }{205} &  \\ \cline{1-4}
\multicolumn{1}{ |c| }{$m_s< T_r < m_c $} & $s\bar{s}$ & \multicolumn{2}{ |c|| }{247} &  \\ \cline{1-4}
\multicolumn{1}{ |c| }{$m_c< T_r < m_\tau $} & $c\bar{c}$ & \multicolumn{2}{ |c|| }{289} &  \\ \cline{1-4}
\multicolumn{1}{ |c| }{$m_\tau< T_r < m_b $} & $\tau^\pm$ & \multicolumn{2}{ |c|| }{303} &  \\ \cline{1-4}
\multicolumn{1}{ |c| }{$m_b< T_r < m_{W, Z} $} & $b\bar{b}$ & \multicolumn{2}{ |c|| }{345} &  \\ \cline{1-4}
\multicolumn{1}{ |c| }{$m_{W, Z}< T_r < m_H $} & $W^\pm, Z^0$ & \multicolumn{2}{ |c|| }{381} &  \\ \cline{1-4}
\multicolumn{1}{ |c| }{$m_H< T_r < m_t $} & $H^0$ & \multicolumn{2}{ |c|| }{385} &  \\ \cline{1-4}
\multicolumn{1}{ |c| }{$m_t < T_r$}  & $t\bar{t}$ & \multicolumn{2}{ |c|| }{427} &  \\ \hline
\end{tabular}}
\end{center}
\caption[]{Effective relativistic degrees of freedom with increasing temperature (adapted from \cite{physreview} \footnotemark[1]).
 }  \label{table:N}
\end{table}
\footnotetext[1]{The difference between \cite{physreview} and our analysis is that we account for both Majorana and Dirac neutrinos and explicitly separate out
the number of degrees of freedom of neutrinos $N_\nu(T_\nu)$ after they decouple from radiation.}

Table \ref{table:N} reflects also the change in $N$ as neutrinos decouple from equilibrium. It is worth noting that
we consider neutrinos to remain relativistic until the present epoch.
Let us first discuss the case of Majorana neutrinos,
disregarding the last column ($N_s(T_s)$). Neutrinos decouple at a
temperature $T_{D_\nu}$ when their interaction rate with electrons becomes too low for equilibrium to be maintained. This means that no annihilation occurs when
$T_{D_\nu}$ is reached and consequently no thermal annihilation energy is available \cite{dodelson}. The degrees of freedom of neutrinos
become separate from the degrees of freedom of radiation when they decouple, but neutrinos continue sharing the temperature of radiation until the next
annihilation occurs, which happens at the electron-positron threshold. If we consider $N=N_r(T_r)+N_\nu(T_\nu)$ after the decoupling threshold,
then in the interval between $T_{D_\nu}$ and $m_e$ we have $T_r=T_\nu$. Once the temperature reaches $m_e$, positrons and electrons annihilate and the
annihilation energy is deposited in radiation but not in the neutrinos. From then on, $T_r > T_\nu$.

Considering now the case of Dirac neutrinos, there exist twice as many neutrino states, as compared to the Majorana case,
since antineutrinos are now distinct from neutrinos. Experiments however only observe left-handed neutrinos, i.e.
neutrinos with negative helicity and antineutrinos with positive helicitiy. Therefore, if the other helicities
do exist we need to consider them to be sterile components. Whereas the active neutrinos have the same degrees of freedom
as in the Majorana case (denoted by the index $\nu$ in Table \ref{table:N}), the sterile neutrinos (index $s$, last column) are not
coupled to the rest of the universe at all, exhibiting a mere $T_s\propto Q^{-1}$ dependence.

From particle thermodynamics, $N(T)$ enters the density of radiation as \cite{physreview}

\begin{equation} \label{eq:rhorad}
\rho = \frac{\pi^2}{30}N(T)T^4,
\end{equation}
and the entropy transfer in an annihilation process (from state $b$ to $a$) allows us to write \cite{KeT}
\begin{equation} \label{eq:NaNb}
N_{b}(QT_b)^3 = N_{a}(QT_a)^3 \implies T_a = \left(\frac{N_b}{N_a}\right)^{\frac{1}{3}}T_b.
\end{equation}
Thus, for the relation between the density before and after the annihilation of a species, in terms of the density parameter, we have
\begin{equation}
\frac{\Omega_a}{\Omega_b}= \frac{N_a}{N_b} \left(\frac{T_a}{T_b}\right)^4 \implies \Omega_a = \left(\frac{N_b}{N_a}\right)^{\frac{1}{3}}\Omega_b.
\end{equation}
To compute the present day density parameter of radiation we use equations (\ref{eq:rhorad}) and (\ref{eq:rho_c})
\begin{equation}
\Omega_\gamma(t_0) = \frac{\rho_\gamma(t_0)}{\rho_c(t_0)}= \frac{\pi^2}{30}N_r(T_{r}(t_0))T_{r}(t_0)^4\frac{8\pi G}{3H_0^2}.
\end{equation}
Knowing that $T_{r}(t_0)=2.725$ K \cite{cmbr}, we obtain
\begin{equation} \label{eq:om_ph}
\Omega_\gamma(t_0)\simeq 4.986\times 10^{-5}.
\end{equation}
Similarly, we can compute the present energy density of Majorana neutrinos using
\begin{equation} \label{eq:om_nu}
\Omega_\nu(t_0) = \frac{\rho_\nu(t_0)}{\rho_c(t_0)}= \frac{\pi^2}{30}N_\nu(T_{\nu}(t_0))T_{\nu}(t_0)^4\frac{8\pi G}{3H_0^2},
\end{equation}
where, from (\ref{eq:NaNb}),
\begin{equation} \label{eq:tnu0}
T_\nu(t_0) = \left(\frac{4}{11}\right)^{\frac{1}{3}}\frac{T_{r}(t_0)}{Q(t_0)} \simeq 1.945 \textsc{ K}.
\end{equation}
According to Table \ref{table:N}, neutrino degrees of freedom are treated separately for times subsequent to $e^\pm$-pair annihilation.
Then we have $N_\nu(T_{\nu}(t_0))=21/4$ and inserting this in (\ref{eq:om_nu}), we obtain
\begin{equation} \label{eq:om_nu0}
\Omega_\nu(t_0) \simeq 3.40 \times 10^{-5}\ .
\end{equation}
We can write the density parameter of radiation as
\begin{equation} \label{eq:cteprop}
\Omega_r(t) = \frac{1}{\zeta}\left[N_r(T_r)T_r(t)^4 + \Theta(T_{D\nu}-T_r)N_\nu(T_\nu)T_\nu(t)^4\right],
\end{equation}
with \begin{equation}
\zeta \equiv \frac{90 H_0^2 }{8 \pi^3 G},
\end{equation}
where, between a certain particle threshold and the next threshold at $t_i$, only $T_r(t)$ changes, while the other terms of $\Omega_r(t)$ remain constant. The
step function $\Theta$ expresses the decoupling of neutrinos below $T_r=T_{D_\nu}$. \\
Writing (\ref{eq:rhorad}) in terms of the radiation temperature and considering that the scale factor relates to it via (\ref{eq:3rad}), we get
\begin{equation} \label{eq:trad}
T_{r}(t)=\left(\zeta \frac{ \Omega_{r}^T(t_i)}{N(t)}\right)^{\frac{1}{4}}\frac{1}{Q(t)},
\end{equation}
where $\Omega_r^T(t)$ is identical to $\Omega_r(t)$ before the $e^\pm$ threshold and only accounts for the photons after the threshold \cite{tese}.\\
In the Dirac neutrino case, the active neutrinos behave as the Majorana neutrinos described above. The sterile neutrinos
follow their own temperature dependence and we assume that they start out at the same temperature $T_1$ as the rest of
the universe at the end of inflation, $t=t_1$. From then on,
\begin{equation}
\Omega_s(t)=\Omega_s(t_1)/Q(t)^4\ ,
\end{equation}
where
\begin{equation}
\Omega_s(t_1)=\zeta^{-1}\frac{21}{4}T_1^4 Q(t_1)^4\ .
\end{equation}

\subsection{Modelling the temperature}

The process of a species leaving equilibrium takes a certain time to occur --- for every particle to stop scattering with the species that remain thermalized, over the whole universe. As the temperature drops with the expansion and reaches an annihilation threshold, we model the annihilation stage by assuming that the energy liberated in the annihilations maintains the temperature constant until all annihilations cease. We define the step function N(t) according to Table \ref{table:N} to describe the change of degrees of freedom available before and after the transitions.
During the transitions, the temperature is kept constant by requiring $N(t)=S(t)$, where
\begin{equation} \label{eq:Smaj}
S(t) = \left(\zeta\frac{\Omega_{r}^T(t_i)}{Q(t_i)Q(t)^3}\right)\times \begin{cases}
      m_e^{-4}, & T_r(t)= m_e\\
      m^{-4}_{\mu}, & T_r(t)= m_{\mu}\\
      m^{-4}_{\pi}, &  T_r(t)= m_{\pi}\\
      T^{-4}_c, &  T_r(t)=  T_c\\
      m^{-4}_s, & T_r(t)= m_s\\
      m^{-4}_c, & T_r(t)= m_c\\
      m^{-4}_{\tau}, & T_r(t)= m_{\tau}\\
      m^{-4}_b, & T_r(t)= m_b\\
      m^{-4}_{W, Z}, & T_r(t)= m_{W, Z}\\
      m^{-4}_H, & T_r(t)= m_H\\
      m^{-4}_t, & T_r(t)= m_t\\
     \end{cases}.
\end{equation}

To model the dark matter temperature, we consider one of the proposed particle candidates as responsible for the observed 23\% of the
density of the universe assigned to dark matter. Since the thermal decoupling of neutralinos has been quantitatively studied in \cite{bringmann} and \cite{hofmann}, we will consider them in our approach. Neutralinos drop from the thermal chemical equilibrium with radiation when the expansion of the universe makes it difficult for a neutralino to find another one to annihilate with. However, by collisions with fermions, neutralinos maintain the kinetic equilibrium and continue sharing the radiation temperature.
The temperature at which neutralinos kinetically decouple if their masses are known is according to \cite{hofmann}
\begin{equation} \label{eq:kindec}
T_{kd} = \left(1.2 \times 10^{-2} \frac{m_{Pl}}{M_{\tilde{\chi}}(M_{\tilde{L}}^2 - M_{\tilde{\chi}}^2)^2}\right)^{-\frac{1}{4}},
\end{equation}
where $m_{Pl}$ is the Planck mass, $M_{\tilde{L}}$ the mass of the sfermions (that in this approach is assumed to be
the same for all sfermions) and $M_{\tilde{\chi}}$ the mass of the neutralino.
Knowing that $T_{DM}(t_0) = T_{kd}(M_{\bar{\chi}})Q_{kd}^2$, we can write a step function for dark matter, that will be
identical to (\ref{eq:trad}) until neutralinos decouple kinetically,
when they are non-relativistic and follow (\ref{eq:mat3}) afterwards
\begin{equation} \label{eq:dmtemp}
T_{DM}(t) = \begin{cases}
 T_r(t), & t < t_{kd}(M_{\bar{\chi}}) \\
 T_{kd}(M_{\bar{\chi}})Q_{kd}^2Q(t)^{-2}, & t \geq t_{kd}(M_{\bar{\chi}})
     \end{cases}.
\end{equation}

\subsection{Equation of motion}

We are now ready to write the equation of motion for this universe, by rearranging Friedmann's equation (\ref{eq:00}) and writing the density parameter of each entity as a function of the scale factor, obtaining
\begin{equation}\label{eq:Q(t)ponto}
\dot{Q}(t)= \sqrt{\frac{\Omega_{r}(t_i)+\Omega_s(t_1)}{Q(t)^2}  + \frac{\Omega_b(t_0) + \Omega_c(t_0)}{Q(t)} +
\Omega_k(t_0) + \Omega_{\Lambda}Q(t)^2}\ ,
\end{equation}
where the sterile neutrinos are included if $\Omega_s(t_1)$ is not set to zero.

Solving equation (\ref{eq:Q(t)ponto}) numerically, using the Fehlberg fourth-fifth order Runge-Kutta method, we obtain the scale factor as a function of time and it allows us to compute the temperature and density parameter. The 7-year WMAP observations \cite{WMAP7} and eqs. (\ref{eq:om_ph}) and (\ref{eq:om_nu0})
determine the present-epoch values of the density parameters of each component. For radiation
\begin{equation}
\Omega_r(t_0)= \Omega_\gamma(t_0)+\Omega_\nu(t_0)\ .
\end{equation}
For the present density parameter of matter, we use
\begin{equation}
\Omega_{mat}(t_0)=\Omega_b(t_0) + \Omega_{c}(t_0).
\end{equation}
The time $t$ will be given in units of the Hubble time $H_0^{-1}$ and the initial time and scale factor values correspond
roughly to the estimates of the end of inflation ($t \sim 10^{-32} \text{s}$ and $Q(t) \sim 10^{-27}$) \cite{KeT}.

\section{Results}

\subsection{Scale factor}

\begin{figure}[h!]
\centering
\includegraphics[width=.7\textwidth]{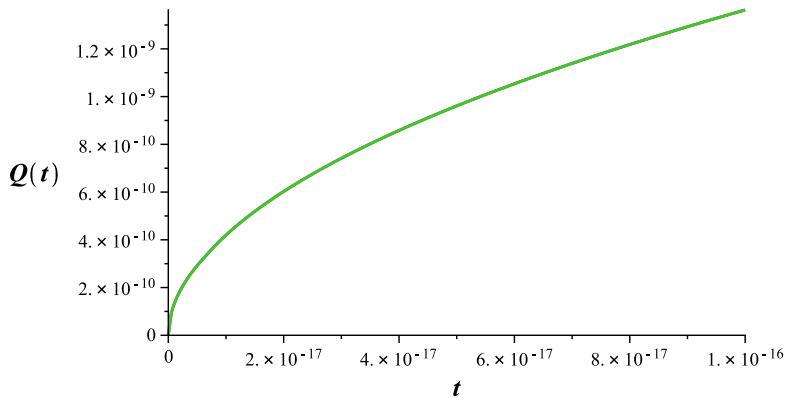}
   \caption{Scale factor over the range of particle annihilations as function of time, with $t$ in $H_0^{-1}$ units.} \label{pic:QMQD}
\end{figure}

The proportionality constant of the radiation density parameter varies at each particle annihilation threshold due to
the decrease in degrees of freedom
during the expansion of the universe. We can solve the equation of motion with a different constant for each time
interval between two annihilation thresholds. This results in a set of 12 solutions.
In terms of temperature, the particle annihilation thresholds do not present a
very significant deviation from the law $T \propto Q^{-1}$ except in the $e^\pm$ case.
The scale factor computed from just two differential equations, one to describe the universe before the annihilation
of electrons and positrons, and another to describe it afterwards, yields a very good approximation \cite{tese}, but
the cumulative effect of all the annihilation processes is noticeable when considering the temperature evolution.\\
A feature we investigated was how the scale factor modifies if neutrinos are Dirac or Majorana particles.
It turns out that the overall results are almost indistinguishable in both cases. In Figure \ref{pic:QMQD}
the evolution of the scale factor in the time range where the particle annihilation processes occur is shown.
\\

The age of the universe is model dependent --- a universe with a dark energy component is older than one comprised
only of matter and radiation, since the expansion rate is accelerating in a model with $\Lambda$  ---
and in our study, which includes dark energy, the value that $t$ attains when $Q_0 =1$ is $t_0 \simeq 0.9983$ $H_0^{-1}$,
equivalent to 13.87 Gyrs.

\subsection{Temperature}

\subsubsection{Radiation Temperature}

\begin{figure}[h!]
\begin{center}
\includegraphics[width=.73\textwidth]{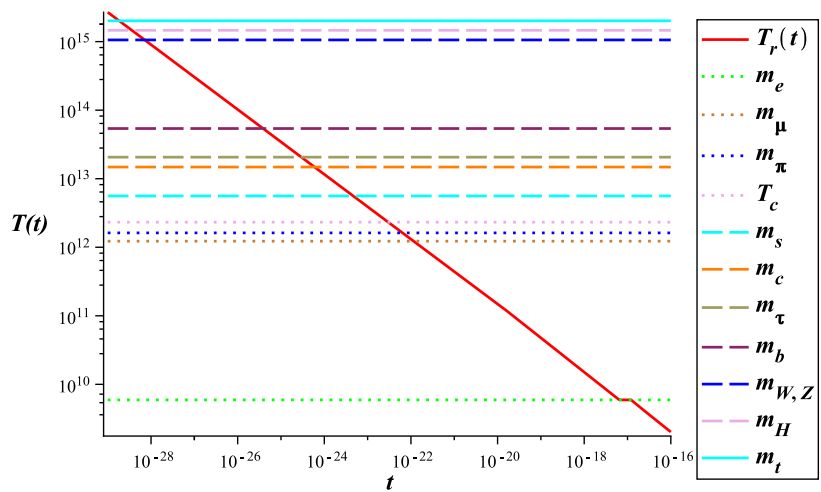}
\end{center}
\caption{Radiation temperature in the range of particle annihilation vs. time, with $t$ in $H_0^{-1}$ units and $T$ in K.} \label{pic:Tstepover}
\end{figure}

\begin{figure}[tbp]
\begin{minipage}{0.35\textwidth}
\includegraphics[scale=0.25]{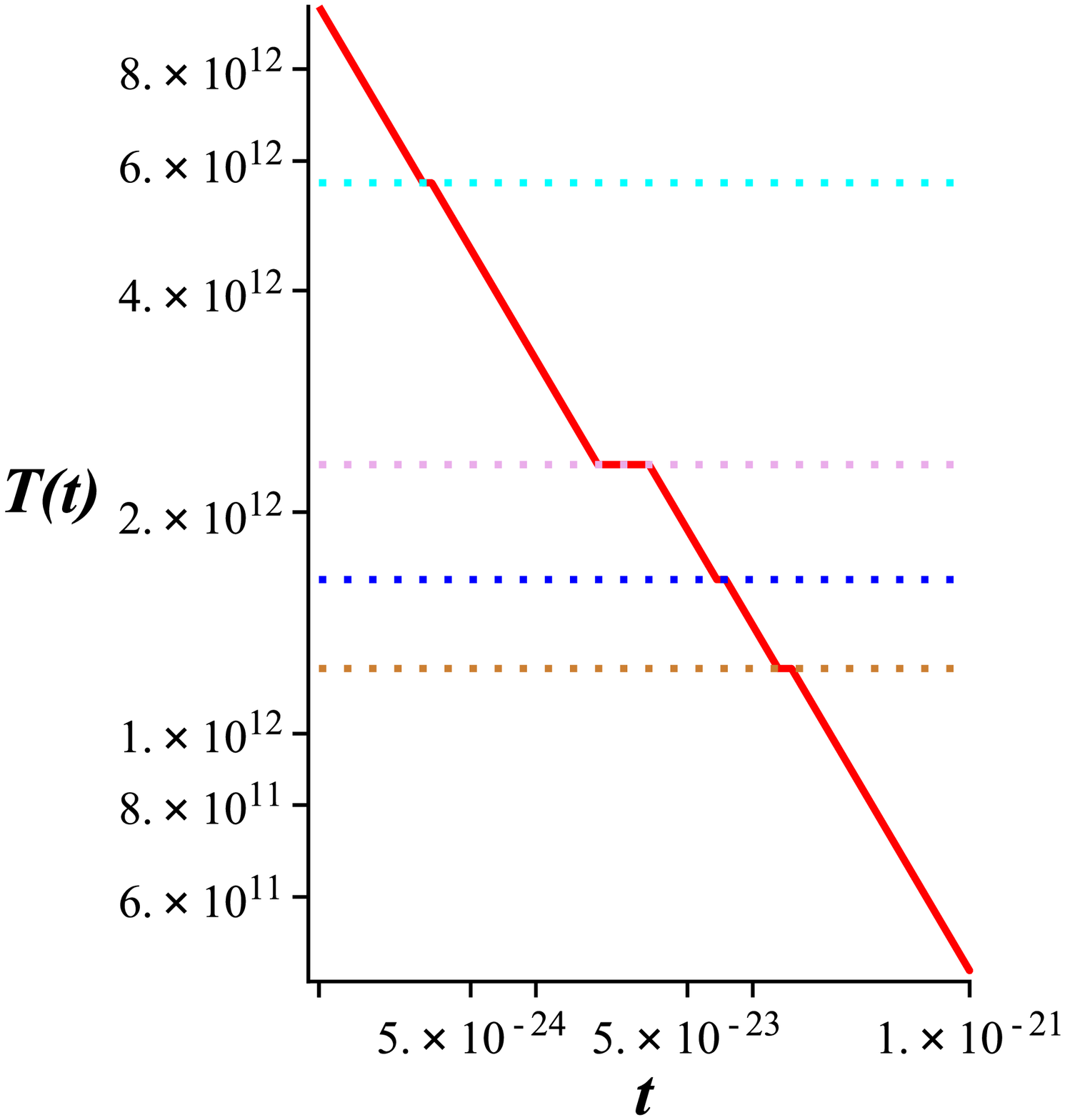}
\end{minipage}
\begin{minipage}{0.3\textwidth}
\includegraphics[scale=0.22]{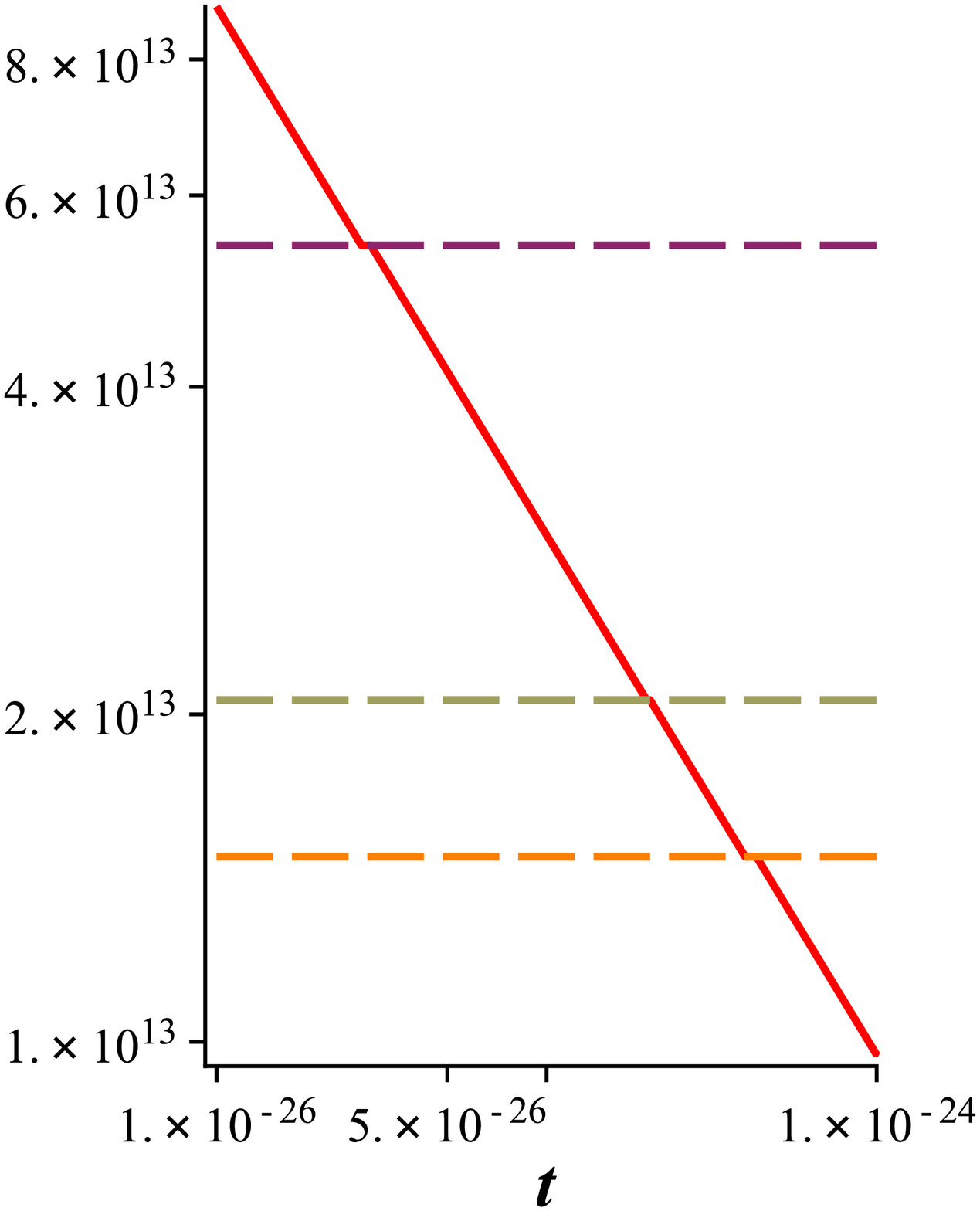}
\end{minipage}
\begin{minipage}{0.3\textwidth}
\includegraphics[scale=0.22]{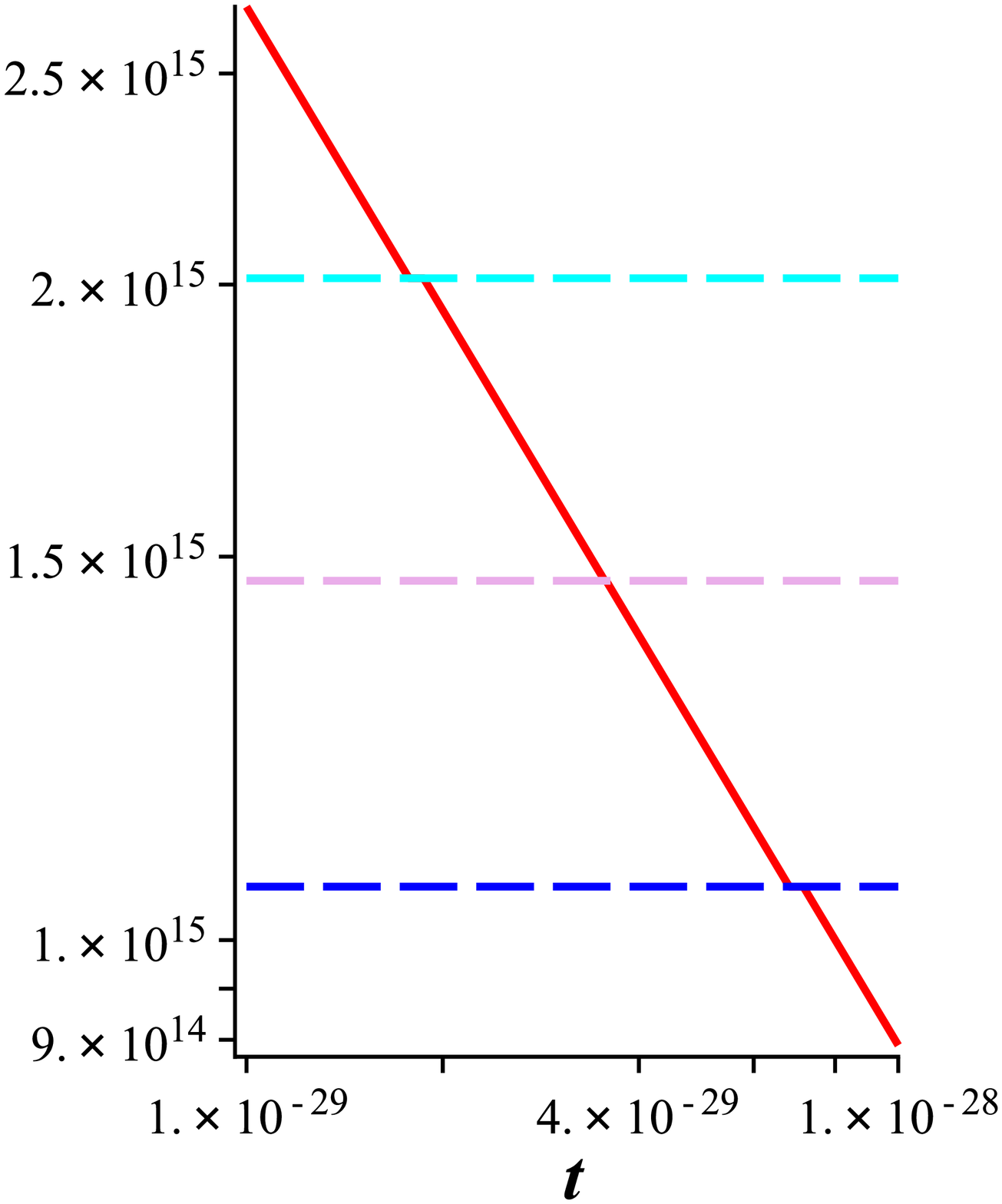}
\end{minipage}
\begin{minipage}{0.01\textwidth}
\includegraphics[scale=0.4]{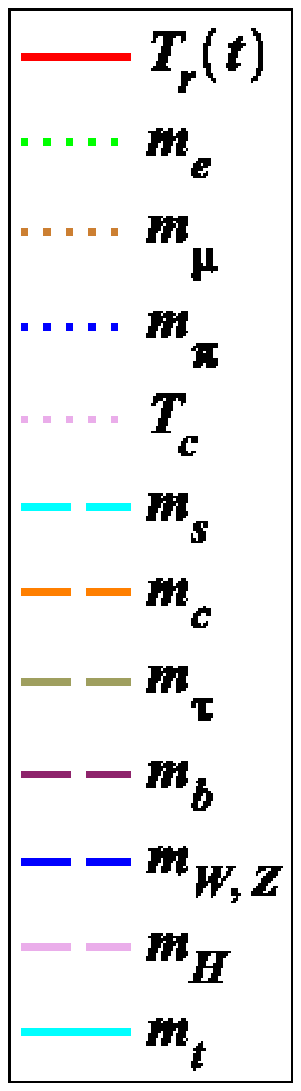}
\end{minipage}
   \caption{Radiation temperature vs. time, with $t$ in $H_0^{-1}$ units and $T$ in K, showing the threshold of annihilation of (left) $W$ and $Z$ bosons,
Higgs boson and top quark, (center) charm quark, $\tau$ particle and bottom quark, and (right) $\mu$, $\pi$, hadrons and strange quark.} \label{pic:Tstep1}
\end{figure}

Figure \ref{pic:Tstepover} presents the temperature of radiation over the time range which encompasses all the
particle annihilation processes. The step-like temperature behaviour due to the values of $N(t)$ is almost invisible
except at the $e^\pm$ threshold, around $t \sim 10^{-17} H_0^{-1}$. Zooming into the regions where the temperature
equals the rest-mass of the other particle species, the annihilations become noticeable, see Figure \ref{pic:Tstep1}.\\

\subsubsection{Neutrino Temperature}

\begin{figure}[h!]
\begin{center}
\begin{minipage}{0.45\textwidth}
\hspace*{-0.4cm}\includegraphics[scale=0.3]{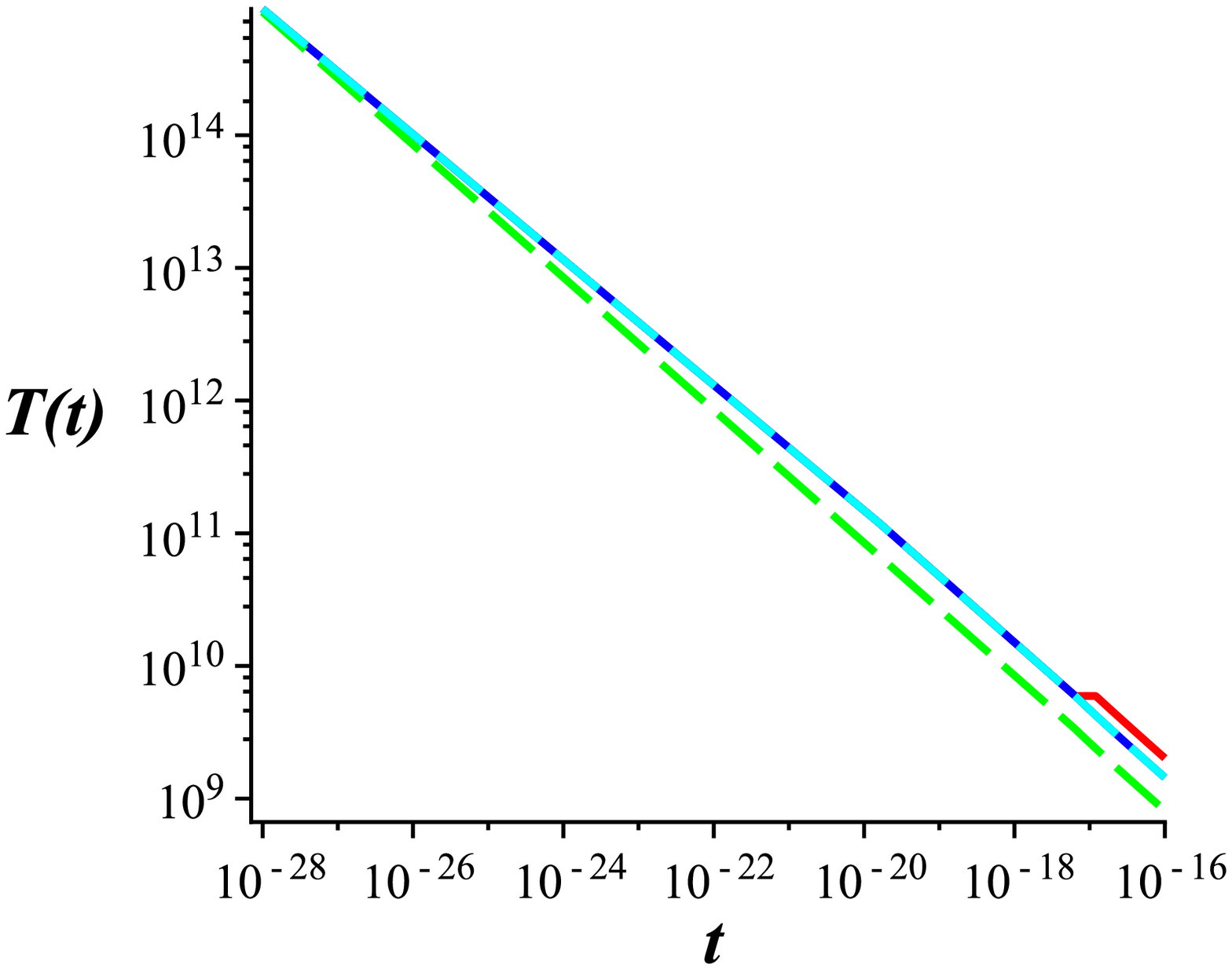}
\end{minipage}
\begin{minipage}{0.4\textwidth}
\hspace*{-1.4cm}\includegraphics[scale=0.3]{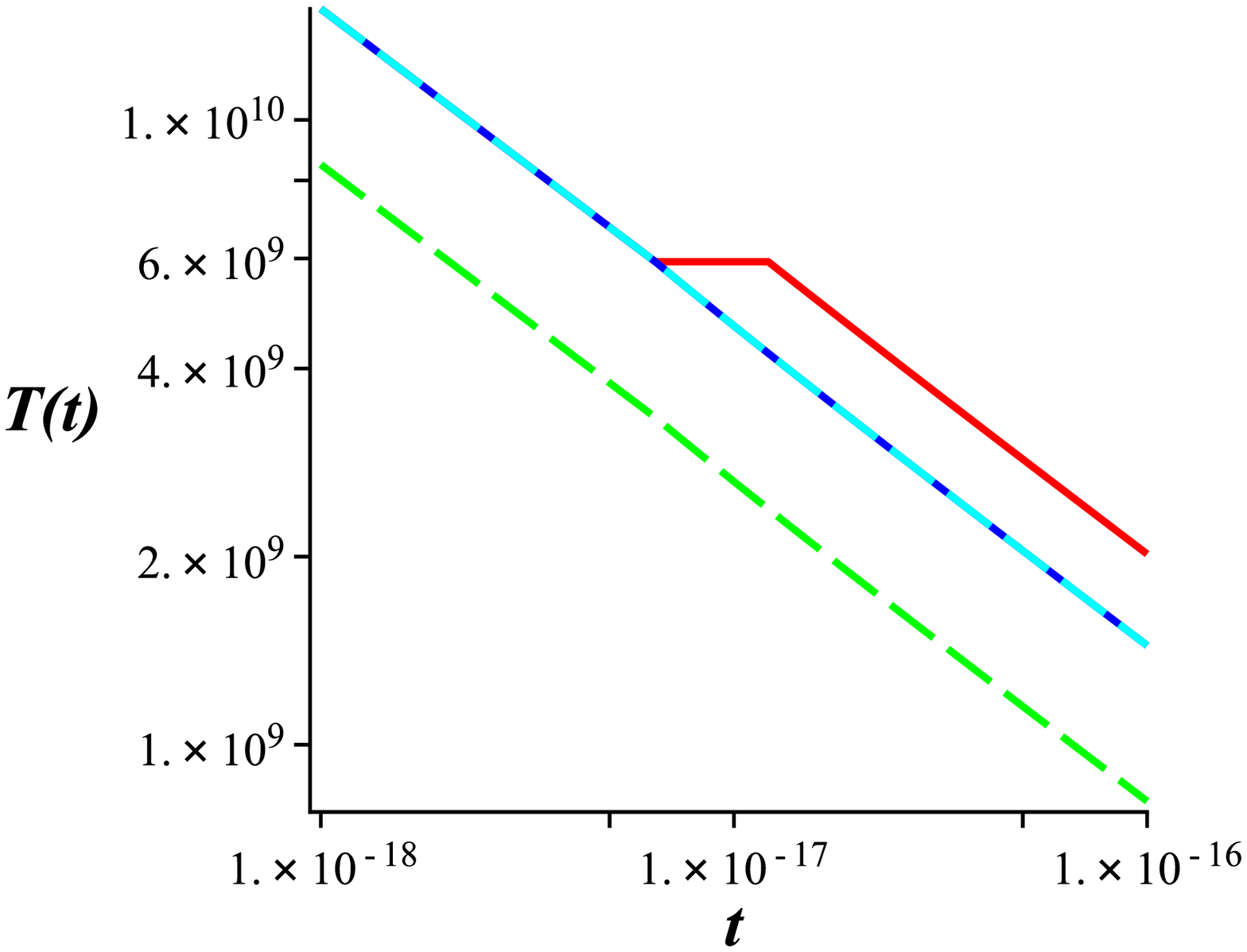}
\end{minipage}
\begin{minipage}{0.01\textwidth}
\hspace*{-1.4cm}\includegraphics[scale=0.5]{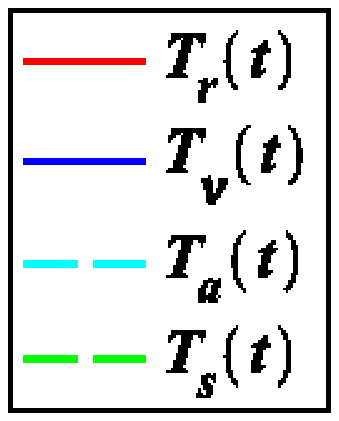}
\end{minipage}
\end{center}
\caption{Radiation temperature $T_r$ and neutrino temperature $T_{\nu}$ in the Majorana case, as well as
the active and sterile neutrino temperatures in the Dirac case, $T_{a}$ and $T_{s}$, as
functions of time, with $t$ in $H_0^{-1}$ units and $T$ in K. On the right, we focus on the
electron-positron annihilation threshold.} \label{pic:TnuTr}
\end{figure}

Majorana neutrinos (or active neutrinos in the Dirac case) differ from photons in temperature because they decouple
shortly before electrons and positrons annihilate and are not sensitive to the associated change in effective degrees
of freedom. In this way, to
describe the neutrino temperature we use the radiation temperature until $t = t_e$, the point at which we alter $N(t)$
to $N_{\nu}$ and $\Omega_{r}^{T}(t_i)$ to $\Omega_{\nu}(t_0)$. Thus, we can write
\begin{equation} \label{eq:tnu1}
T_{\nu}(t)=\left(\zeta \frac{\Omega_{\nu}(t_0)}{N_\nu(t)}\right)^{\frac{1}{4}}\frac{1}{Q(t)}, \hspace{0.5 cm} t \geq t_e,
\end{equation}
while before the time $t_e$ they share the radiation temperature, (\ref{eq:trad}). Plotting
$T_{r}(t)$ and $T_{\nu}(t)$ given by (\ref{eq:tnu1}) in Figure \ref{pic:TnuTr}, we can see the
moment when the active or Majorana neutrinos decouple from radiation, around $t \simeq 10^{-17} H_0^{-1}$. We observe
that having $\Omega_{r}(t_e)$ and $4N = 43$ in (\ref{eq:trad}) is equivalent to having $\Omega_{\nu}(t_0)$ and
$4N = 21$ in
(\ref{eq:tnu1}), because the temperature of radiation before $m_e$ is reached coincides with the temperature of the
active neutrinos when they become a separate species, at $t = t_{T_{D\nu}}$. If neutrinos are of Dirac type, the active
neutrinos behave as in the Majorana case. Since the scale factor does not change noticeably when including sterile
neutrinos, the temperature behavior of active neutrinos $T_a$ is practically indistinguishable from the Majorana neutrino temperature
$T_{\nu}$. The sterile neutrino temperature $T_{s}$ decreases monotonically, unaffected by the
threshold processes.

\subsubsection{Matter Temperature}

\begin{figure}[h!]
\begin{center}
   \includegraphics[width=.6\textwidth]{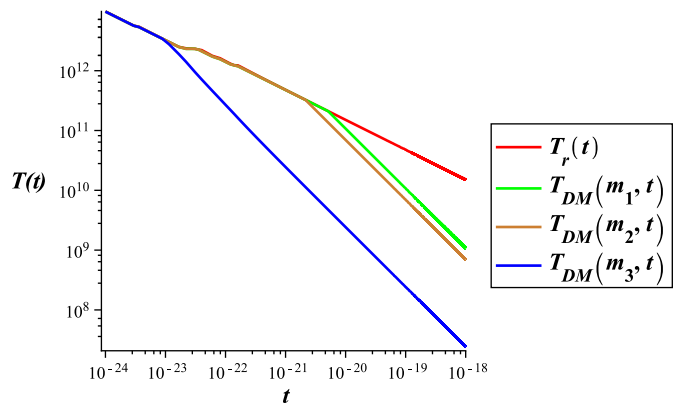}
\end{center}
   \caption{Representation of neutralinos decoupling kinetically from radiation by plotting $T_r(t)$ and $T_{DM}(t)$, with $t$ in $H_0^{-1}$ units and $T$ in K.} \label{pic:Tdmdec}
\end{figure}

For neutralinos, we take three different mass values in reasonable limits, namely $m_1 =$ 10 GeV, $m_2 =$ 100 GeV
and $m_3 =$ 1 TeV and use equation (\ref{eq:dmtemp}) to determine the time when the neutralinos decouple from
thermal equilibrium with radiation, as is depicted in Figure \ref{pic:Tdmdec}. For subsequent times, they are
non-relativistic and their temperature follows the behaviour of the matter temperature ($\propto Q(t)^{-2}$).
At $t_0$ their temperatures yield $T(m_1) \simeq 1.79 \times 10^{-11}$ K, $T(m_2) \simeq 1.16 \times 10^{-11}$ K and $T(m_3) \simeq 4.70\times 10^{-13}$ K. \\
The baryonic component of matter, created around $ T_c \simeq 2.3 \times 10^{12}$ K at the quark-hadron confinement, can already be considered non-relativistic from the time it was formed. Baryons remained in thermal contact with radiation until decoupling,
which according to estimates \cite{physreview} happened around the temperature
$T_D \simeq 3000$ K, corresponding in this work to a universe aged 373 thousand years,  consistent with other works (e.g. $376\pm 4.1$ thousand years cf. \cite{WMAP9}). At that time the radiation and baryonic matter
temperatures were still the same. This translates as
\begin{equation}
T_D=T_r(t_D)=\frac{T_r(t_0)}{Q_D} \implies Q_D = \frac{T_{r}(t_0)}{T_D} \simeq 9.083 \times 10^{-4},
\end{equation}
and
\begin{equation}
T_D=T_{mat}(t_D)=\frac{T_{mat}(t_0)}{Q_D^2} \implies T_{mat}(t_0)=T_DQ_D^2 \simeq 2.475 \times 10^{-3} \text{K}.
\end{equation}
The present epoch temperatures of baryonic and dark matter are idealized averages of the matter temperatures in the universe. Since baryonic matter and most probably dark matter clustered into structures, matter-rich regions are expected to possess much higher temperatures than low-density space.

\subsection{Density}

\begin{figure}[h!]
\begin{center}
   \includegraphics[width=.7\textwidth]{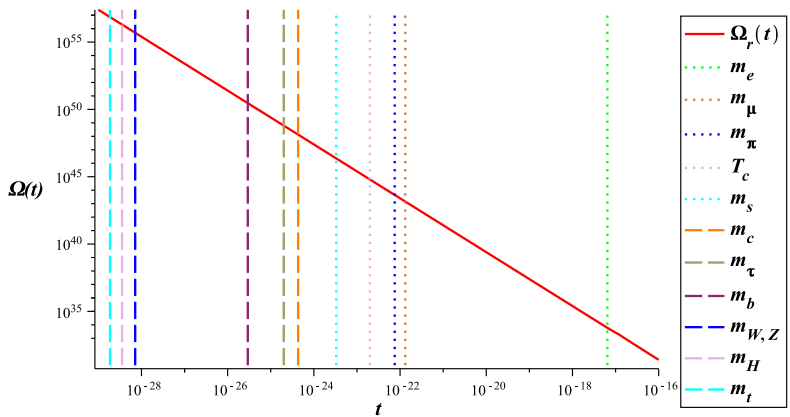}
\end{center}
   \caption{Radiation density parameter in the range of particle annihilations, with $t$ in $H_0^{-1}$ units.} \label{pic:Om2eOm12}
\end{figure}

\begin{figure}[h!]
\begin{center}
   \includegraphics[width=.7\textwidth]{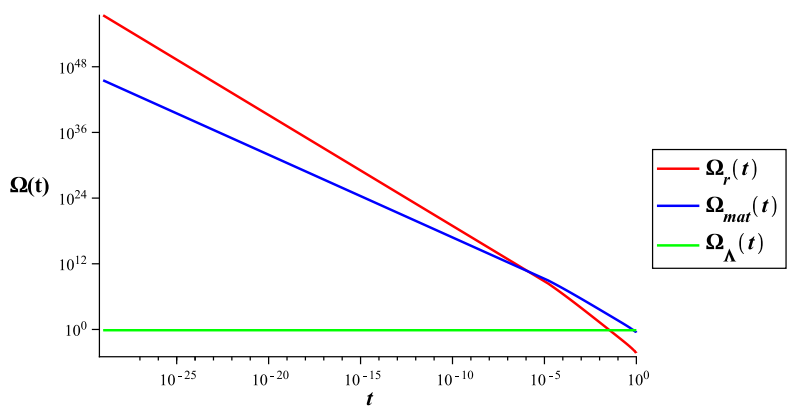}
\end{center}
   \caption{Representation of the density parameters over time, with $t$ in $H_0^{-1}$ units, from the beginning of the particle thresholds to the present time.} \label{pic:Ooverview}
\end{figure}

The radiation density parameter includes photons and neutrinos (which we assume as Majorana throughout this section) and is given by (\ref{eq:cteprop}).
Plotting the radiation density in Figure \ref{pic:Om2eOm12}, we note that at this scale the annihilation processes have no visible effect.
If we zoomed into the regions where the various species annihilate, as we did in the case of the radiation temperature, we would observe a small effect at each stage.
\\
Dark energy vs. time is a constant, $\Omega_\Lambda$, while matter density is divided into the dark matter and baryonic parts. The latter appears only after baryogenesis, when part of the radiation density is transferred to the new-born baryons. We can describe this process using again a step function
\begin{equation}
\Omega_{mat}(t) = \begin{cases}
\Omega_{c}(t_0)Q(t)^{-3}, & t < t_{T_c} \\
 \left[\Omega_{c}(t_0)+\Omega_{b}(t_0)\right]Q(t)^{-3}, & t \geq t_{T_c}
     \end{cases}.
\end{equation}
Plotting an overview of the behaviour of the density contributions in a general universe in Figure \ref{pic:Ooverview}, we see that the matter density decreases more slowly than the radiation density, leading to a matter-radiation equality point. The intersection between the two components happens in the simulation at $t \simeq 1.6 \times 10 ^{-6} H_0^{-1} \simeq 22$ thousand years, when the universe becomes matter-dominated, thus the universe was already matter-dominated when baryonic matter decoupled from radiation. We observe the dark energy component, constant over time, gaining importance for later times and the point when it becomes the dominant component of the universe, at $t = 0.69 H_0^{-1}$, corresponds to 9.6 Gyrs.

Let us now focus on the neutrino densities. The number of neutrino species affects primordial nucleosynthesis
and according to {\cite{Dolgov2002}} sterile neutrinos should be suppressed by a factor $\simeq 0.05$ at the epoch of
$T=1$ MeV and can never have been in thermal equilibrium if they are light (masses below 30 eV). It is interesting to
observe that our calculation, which starts at the end of inflation assuming a {\sl common} initial temperature for
all species, attains the value $\rho_s/\rho_\nu=0.047$ at the epoch of nucleosynthesis. The suppression factor stems from
the lower temperature of the sterile neutrinos at nucleosynthesis, since they did not acquire any of the annihilation
energies which the active neutrinos accumulated until then.

\section{Conclusions}

Through modelling the universe as a perfect fluid in terms of $Q(t)$, $\Omega(t)$ and $T(t)$, interpreting dark energy
as arising from fluctuations of the metric and assuming neutralinos as dark matter particles, we could reconstruct many
of the important events of its evolution. Starting from the known present epoch parameters, we show the temperature and
density effects of particle annihilation and estimate the average present temperature of dark matter if it was composed
of neutralinos.\\
We consider the two possibilities of Majorana and Dirac neutrinos. The overall development of the universe is not
noticeably affected by the extra degrees of freedom of the Dirac neutrinos as compared to the Majorana case. We find the
interesting result that at the epoch of primordial nucleosynthesis the sterile neutrino density, compared to the active
neutrino density, is suppressed by a factor which is in agreement with estimates in the literature, when the sterile
neutrinos start out, at the end of inflation, with the same temperature as the rest of the universe.

\section*{Acknowledgments}

This work was supported in part by Fundação para a Ciência e a Tecnologia, CERN grant no. FP/116334/2010, the 7th Framework Programme grant no. 283286,
and by the "Helmholtz Alliance for Astroparticle Phyics (HAP)" funded by the Initiative and Networking Fund of the Helmholtz Association. AB gratefully
acknowledges a fruitful discussion with A.D. Dolgov about sterile neutrinos.

\bibliographystyle{unsrt}
\bibliography{mybibliography}

\end{document}